\pdfoutput=1

\documentclass[11pt]{article}

\usepackage[]{acl}

\usepackage{times}
\usepackage{latexsym}

\usepackage[T1]{fontenc}

\usepackage[utf8]{inputenc}

\usepackage{microtype}
\usepackage{hyperref}       
\usepackage{url}            
\usepackage{booktabs}       
\usepackage{nicefrac}       
\usepackage{multirow}
\usepackage{enumerate}
\usepackage{subcaption}
\usepackage{graphicx}
\usepackage{xspace}
\usepackage{paralist}
\usepackage{makecell}
\usepackage{amsmath}
\usepackage{tabularx,adjustbox}
\usepackage{todonotes}
\usepackage{breqn}
\usepackage{color,cite}
\usepackage{amssymb}
\usepackage{amsfonts}
\usepackage{bbm}

\usepackage{amsmath,amsfonts,bm}

\def\1{\bm{1}}

\def\vn{{\bm{n}}}

\DeclareMathAlphabet{\mathsfit}{\encodingdefault}{\sfdefault}{m}{sl}
\SetMathAlphabet{\mathsfit}{bold}{\encodingdefault}{\sfdefault}{bx}{n}

\newcommand{\imdb}{\textbf{IMDB}\xspace}
\newcommand{\sst}{\textbf{SST}\xspace}
\newcommand{\fst}{\textbf{FST}\xspace}
\newcommand{\googlesentiment}{\textbf{Google Cloud}\xspace}
\newcommand{\ibmsentiment}{\textbf{IBM Cloud}\xspace}

\newcommand{\googletranslation}{\textbf{Google Trans}\xspace}
\newcommand{\bingtranslation}{\textbf{Bing Trans}\xspace}

\title{Student Surpasses Teacher: Imitation Attack for Black-Box NLP APIs}

\author{Qiongkai Xu\thanks{Most of the work was finished when the first author was at ANU and Data61 CSIRO.},$^{1}$ Xuanli He,$^{2}$ Lingjuan Lyu,$^{3}$ Lizhen Qu,$^{2}$  Gholamreza Haffari$^{2}$\\
  $^1$The University of Melbourne,
  $^2$Monash University,
  $^3$Sony AI \\
  \texttt{Qiongkai.Xu@unimelb.edu.au, Lingjuan.Lv@sony.com}\\ \texttt{\{xuanli.he1,Lizhen.Qu,Gholamreza.Haffari\}@monash.edu} \\}

\begin{document}
\maketitle

\begin{abstract}

Machine-learning-as-a-service (MLaaS) has attracted millions of users to their splendid large-scale models. Although published as black-box APIs, the valuable models behind these services are still vulnerable to imitation attacks. Recently, a series of works have demonstrated that attackers manage to steal or extract the victim models. Nonetheless, none of the previous stolen models can outperform the original black-box APIs. In this work, we conduct unsupervised domain adaptation and multi-victim ensemble to  showing that attackers could potentially surpass victims, which is beyond previous understanding of model extraction. Extensive experiments on both benchmark datasets and real-world APIs validate that the imitators can succeed in outperforming the original black-box models on transferred domains. We consider our work as a milestone in the research of imitation attack, especially on NLP APIs, as the superior performance could influence the defense or even publishing strategy of API providers.

\end{abstract}

\section{Introduction}
\label{sec:introduction}

Task-oriented NLP APIs have received tremendous success, partly due to commercial cloud services~\citep{krishna2019thieves,wallace2020imitation}. 
The enormous commercial benefit allures other companies or individual users to extract the back-end decision models of these successful APIs.
Some recent works have demonstrated that many existing NLP APIs can be locally imitated or extracted~\citep{krishna2019thieves,wallace2020imitation,guo2022threats}, violating the intellectual property (IP) of NLP APIs. 
Equipped with recent advanced models pre-trained on large-scale corpus, it is getting easier to train a decent attack model with limited training samples retrieved from victims~\citep{he2021model}.

Despite their success, there are still a series of restrictions on current attack paradigm.
The first restriction is that attackers and victims are trained and evaluated under the same domain. Although such setting simplifies the comparison of utility and fidelity between victim and attack models, it is less likely to be the case in real world. Generally, attackers and victims do not, at least not willing to, share their in-house datasets to the public. Moreover, the competition between attackers and victims lies in their performance on the application scenarios (transferred domains), which vary by customers not API providers. 
The second restriction is that traditional model extraction only refers a single victim model, and none of the previous works manage to build the extracted models that can surpass the original black-box APIs~\citep{tramer2016stealing,krishna2019thieves,wallace2020imitation}. Aggregating the strength from diverse victims, however, can potentially benefit the performance of attackers. This may further enable the attack model to acquire a superior performance, in the best case, even surpassing all the corresponding victim models.

Based on the above analysis, we are motivated to address their restrictions by \textit{i)} conducting imitation attacks on a transferred domain and \textit{ii)} utilizing multiple victim ensemble. Our approach integrates unsupervised domain adaptation and model ensemble into the imitation attack. We conduct experiments on two representative NLP tasks, namely sentiment classification~\citep{lyu2020differentially} and machine translation~\citep{wallace2020imitation}.
We investigate the attack performance on both the locally simulated victim models and publicly available commercial NLP APIs. Our results demonstrate that the attackers could potentially achieve better performance than victims in the transferred domains and utilizing multiple victim models further improves the performance of the attack models.
For those target domains that are relatively far from victims' source domains, \textit{e.g.}, sentiment analysis models on movie review domain to financial document domain, the performance improvement of the imitation model could be further amplified.


Overall, our empirical findings exacerbate the potential risk concerns of public API services, as malicious companies could provide a better service in their specific domains by integrating several publicly available APIs. Moreover, the new services generally could cost far less than any of the original victim services or the wage paid to human annotators. The imitation attack not only infringes the IP of victim companies by misusing the predictions of their APIs, but also potentially corrupts the MLaaS market by publishing new APIs, with higher performance but lower price. 
We believe that explicitly exposing the new imitation attack paradigm would arouse significant attentions in the research community and encourage companies to reconsider their strategies of publishing API services. 

\section{Related Work}
\subsection{Model Imitation Attack} 
Model imitation attack (also referred to as model ``extraction'' or ``stealing'') has been studied for simple classification tasks~\citep{tramer2016stealing}, vision tasks~\citep{orekondy2019knockoff}, NLP tasks~\citep{krishna2019thieves, wallace2020imitation}, and \textit{etc}. Generally, model imitation attack aims to reconstruct a local copy or to steal the functionality of a black-box API.
If the reconstruction is successful, the attacker has effectively stolen the intellectual property. 
Previous works on imitation attack~\citep{krishna2019thieves,wallace2020imitation} mainly focus on how to imitate a model with performance approximate to the victim API in the source domain. Whether a more powerful attacker can steal a model that is better than any victim API in new domains is largely unexplored.

\subsection{Unsupervised Domain Adaption}
Domain adaptation is a task of adapting a pre-trained model on a source domain to a target domain. It can be fulfilled in two mechanisms: supervised adaptation and unsupervised adaptation. The former can achieve an outstanding performance with a small amount of in-domain labeled data~\citep{daume-iii-2007-frustratingly,ben2007analysis}. 
In contrast, the unsupervised domain adaptation (UDA)~\citep{miller2019simplified,ganin2015unsupervised} does not require ground-truth labels in the target domain, hence is more challenging but attractive.
 
This work is under the umbrella of UDA family~\citep{wang2021generalizing,miller2019simplified}. We differentiate our work from other UDA works in terms of the intent. Other UDA works aim to improve the models in both source domains and target domains simultaneously, while our proposed imitation attack mainly focuses on optimizing the adapted attack models merely in target domains. We exploit UDA from a dark side, where one can leverage domain adaptation to violate IP of commercial APIs, but benefit from such violations. 

\begin{figure*}
    \centering
    \includegraphics[width=0.92\textwidth]{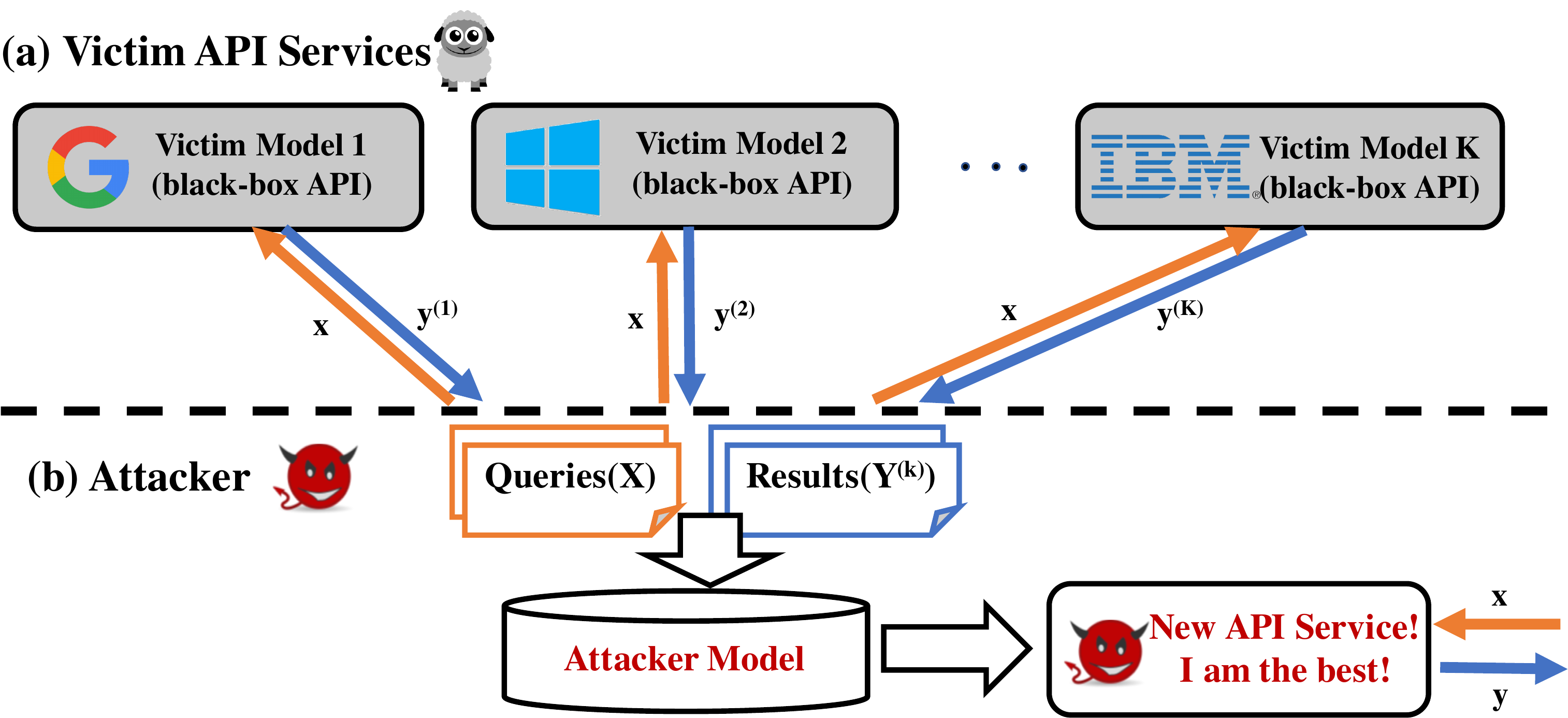}
    \vspace{-2mm}
    \caption{The workflow of the imitation attack and its harm. The attacker labels queries (x) using multiple victim APIs, and then trains an attack model on the resulting data. Finally, the attacker could publish a new API service, which could become a competitor of the victim services, thus infringing intellectual properties and eroding MLaaS market shares of the victim services.}
    \label{fig:attack_pipeline}
    \vspace{-2mm}
\end{figure*}

\subsection{Ensemble for Knowledge Distillation}
Model imitation attack is related to knowledge distillation (KD)~\citep{hinton2015distilling}.
Emergent knowledge distillation~\citep{bucilua2006model,hinton2015distilling} aims to transfer knowledge from a large teacher model to a small student model for the sake of model compression. By encouraging the student model to approximate the behaviors of the teacher model, the student is able to imitate the teacher's behavior with minor quality loss but more efficient inference~\citep{furlanello2018born,dvornik2019diversity,anil2018large}.

However, model imitation differs from distillation in terms of data usage. Particularly, the victim's (\textit{i.e.}, teacher's) training data is usually unknown as the victim API is deployed in a black-box manner. Thus, malicious users tend to consider unlabeled queries more likely to be collected from other domains~\citep{xie2020self} or generated from pre-trained generative models~\citep{ravuri2019classification,kumar2020data,9053146}. Our work considers a more realistic setting, as victim and attack models are under different domains and express distinctive interests.

On the other hand, with the development of multi-task learning, multi-teacher distillation has been proposed, which is targeted at distilling multiple single-task models into a single multi-task model~\citep{tan2019multilingual,clark2019bam,saleh2020collective}. These lines of works lie in the spirit of improving model performance and encouraging parameter reduction via knowledge distillation.

The dichotomy between our work and previous works is we are interested in an ensemble distillation of single-task models, where all teachers and the student work on the same task, but could be in different domain. The gist of this approach aims to leverage the collective wisdom to obtain a better student outperforming its teachers, which is similar to the ensemble learning~\citep{opitz1999popular}. However, ensemble learning focuses more on the aggregation of the predictions of different models at the inference stage, whereas our proposed method not only gathers the inputs and their predictions, but also trains a new model on the input-prediction pairs.

\section{Imitation Attack Paradigm}
In this section, we first explain the problem and motivation of imitation attack (IMA) under domain adaptation setting. Then, we connect our IMA practice with a corresponding domain adaptation theory~\citep{ben2010theory,wang2021generalizing}. After that, we introduce a multi-victim ensemble methodology for IMA. Finally, we explain the rationality of a family of existing defense technologies under the domain adaptation theory.

\subsection{Problem Statement}
\label{sec:problem_statement}
In the real world, attackers may be interested in their own new business, \textit{e.g.}, a new classification or machine translation system in a new domain. Generally, the attackers possess their own training samples, $\mathcal{X}=\{x_1, x_2, \cdots, x_N\}$, while the oracle labels $\mathcal{Y}=\{y_1, y_2, \cdots, y_N\}$ of these samples are not yet labeled. 
In order to train a model with the least cost of annotation, attackers access the publicly available commercial APIs for the target task $\mathcal{T}$. Moreover, the attackers could inquire multiple APIs for further performance improvement. The underlying models of the attacked APIs are $K$ victim models, $\mathcal{V}=\{h_v^{(1)}, h_v^{(2)}, \cdots, h_v^{(K)}\}$. As illustrated in Figure~\ref{fig:attack_pipeline}, our attack can be formulated as a two-step process: 
\begin{enumerate}
\item The attacker queries the $k$-th victim model $h_v^{(k)}$ with samples in $\mathcal{X}$, and retrieves corresponding labels $\mathcal{Y}^{(k)}=\{h_v^{(k)}(x_1), h_v^{(k)}(x_2), \cdots, h_v^{(k)}(x_N)\}$.

\item The attacker learns an imitation (attack) model $h_a(x)$ based on the queries and concatenated retrieved labels, $\langle \mathcal{X}, \mathcal{Y}^{(k)} \rangle |_{k=1}^K$. 
\end{enumerate}
As the attacking process involves teaching a local model by imitating the behavior of the victim models, we put this attack process under the umbrella of imitation attack.
The motivation of such attack paradigm is twofold:
\begin{itemize}
    \item Firstly, querying commercial APIs generally costs far less than hiring human annotators. Therefore, the price for training the attack models using imitated labels is competitive in the market;
    \item Secondly, the attackers are potentially able to outperform the victims in terms of utility. As demonstrated in Section~\ref{sec:exp_result}, domain adaptation and multi-victim ensemble further enhance the attacker performance.
\end{itemize}
We believe that both price and performance advantages would lure companies or individuals to imitation attack.
\subsection{Training for Imitation Attack}
In imitation attack, the attacker utilizes the labels from victim APIs for model training.
Given a victim model 
 $h_v(x)$, the attacker model $h_a(x)$ imitates the behavior of victim model by minimizing the prediction error on the target domain $\mathcal{T}$,
 \begin{equation}
    \label{eq:attack_loss}
     \min_{h_a} \mathbb{E}_{x \sim \mathcal{T}} \left[\mathcal{L}(h_a(x), h_v(x))\right].
 \end{equation}

On the other hand, we assume that the victim model has learned from oracle annotations $y$ in another source domain $\mathcal{S}$, although those labels are never used for training attacker in our IMA. The loss for training victim model is

 \begin{equation}
    \label{eq:victim_loss}
     \min_{h_v} \mathbb{E}_{x \sim \mathcal{S}} \left[\mathcal{L}(h_v(x), y)\right],
 \end{equation}
where $\mathcal{L}(\cdot,\cdot)$ is the loss function. In practice, we use cross entropy as loss functions for training both the victim and attack models.
Note that jointly optimizing Equation~\ref{eq:attack_loss} and Equation~\ref{eq:victim_loss} can derive the unsupervised domain adaptation loss in \citep{miller2019simplified,ganin2015unsupervised}. However, our approach optimizes victim and attack models in absolutely separate steps, as victim models and their training processes are black-box to the attackers.



\subsection{Imitation Attack as Domain Adaptation} 
We connect our new IMA paradigm with domain adaptation theory. The error of the attacker is measured by attacker risk $\epsilon_{a}(h, h_a) :=  \mathbb{E}_{x \sim \mathcal{T}} [|h(x)-h_a(x)|]$. According to domain adaptation theory~\citep{ben2010theory,wang2021generalizing}, the upper bound of the attacker risk is 
\begin{align}
    \label{eq:ima_da}
    \epsilon_{a}(h, h_a) & \leq \epsilon_{v}(h, h_v) + 2d(P_X^{s}, P_X^{t})\\ \nonumber
    &+ \min_{\mathcal{T} \in \{P_X^{s}, P_X^{t}\}} \mathbb{E}_{x \sim \mathcal{T}}\left[|h_v(x)-h_a(x)|\right],
\end{align}
where $d(P_X^{s}, P_X^{t})$ is the total variation between the distributions of source and target domains, which is a constant determined by domains of the datasets used by the victim and attacker. The first term, $\epsilon_{v}(h, h_v) := \mathbb{E}_{x \sim \mathcal{S}} [|h(x)-h_v(x)|]$, with oracle decisions $y=h(x)$, is the victim risk on the source domain. It is optimized during the training of victim models as Equation~\ref{eq:victim_loss}. The last term is associated with imitation training objective as Equation~\ref{eq:attack_loss}. Therefore, our imitation attack under domain adaptation setting is actually optimizing the upper bound of the attacker risk $\epsilon_{a}(h, h_a)$.

\subsection{Multiple Victim Ensemble}
Another approach to achieving further performance improvement is to integrate the results from multiple APIs. This strategy is well motivated in real-world imitation attack, as many cloud computing companies share similar APIs for main-stream NLP tasks, e.g, Google Cloud and Microsoft Azure both support sentiment classification and machine translation. Attackers can improve their performances by learning from multiple victim APIs. In more detail, given independent $K$ victim models $h_{v}^{(k)}(x)$, attackers can take the average advantage of all victim models by,
\begin{equation}
    \min_{h_a} \frac{1}{K} \sum_{k=1}^{K} \mathbb{E}_{x \sim \mathcal{T}} [\mathcal{L}(h_a(x), h_v^{(k)}(x))].
\end{equation}
According to ensemble theories~\citep{breiman2001random,bauer1999empirical}, the lower generalization error of an ensemble model depends on \textit{i)} better performance of the individual models, and \textit{ii)} lower correlation between them. In real-world, companies are actually \textit{i)} targeting on API with better performance, and \textit{ii)} using their own private training datasets. The effort of these companies towards superior API performance unfortunately exacerbates these two factors for a successful ensemble model as an attacker. 

\subsection{Defending Imitation Attack}
Some existing IMA defense strategies~\citep{he2021model} slightly distort the predictions of victim models to reduce the performance of attackers, by replacing $h_v(x)$ with noisy $h'_v(x)$. The variance introduced by distortion is $\Delta (h_v, h_v') := \mathbb{E}_{x \sim \mathcal{S}} [|h_v(x)-h_v'(x)|]$. As the distortion should not destroy the utility of victim model, the variance should be bounded by a small constant $\delta$, \textit{i.e.}, $\Delta (h_v, h_v') \leq \delta$. The victim risk, the first term in \eqref{eq:ima_da}, is relaxed to $\epsilon_{v}(h, h_v')$, where
\begin{equation}
    \epsilon_{v}(h, h_v') \leq \epsilon_{v}(h, h_v) + \Delta (h_v, h_v').
\end{equation}
Therefore, the gentle distortion of victim outputs results in a more relaxed upper bound for optimization, which could potentially lead to better results on defending the imitation attack.

\section{Experimental Setup}
\label{sec:exp_setup}
\subsection{Tasks and Datasets}
In this paper, we focus on two essential NLP tasks, classification and machine translation, both of which are predicting discrete outputs, and cross entropy loss is used as the objective in optimization. For classification tasks, as APIs provide continuous scores of the predictions, we consider the settings of soft label (using prediction scores) and hard label (using predicted categories).
In machine translation, the translation result is a sequence of discrete words, which are considered as sequential hard labels. Classification and translation tasks are evaluated by accuracy(\%)~\citep{schutze2008introduction} and BLEU~\citep{papineni2002bleu}, respectively. The datasets used in our imitation attack experiment are summarized in Table~\ref{tab:dataset}.



\begin{table*}[t]
    \centering
    \small
    \begin{tabular}{llllll}
      \toprule
    
Dataset &  \#Train & \#Dev& \#Test & Task  & Domain\\
    \midrule
{IMDB} &  25,000 & 25,000 & N/A & Sentiment Classification & Movie Review (long) \\
{SST} &  6,920 & 872 & 1,821 & Sentiment Classification & Movie Review (short)\\
{FST} & 1,413 & 159 & 396 & Sentiment Classification & Finance Document \\
    \midrule
{WMT14}& 4.5M & 3,000 & N/A& Machine Translation & General\\
{JRC-Acquis}&2M&1,000&1,000&Machine Translation & Law\\
{Tanzil}&579k&1,000&1,000&Machine Translation& Koran\\
    \bottomrule
 \end{tabular}
\vspace{-2mm}
    \caption{Statistic of sentiment classification and machine translation datasets, with number of samples in train, dev and test sets. The task name and corresponding domain for each dataset are also included.}
    \vspace{-2mm}
    \label{tab:dataset}
\end{table*}


\paragraph{Sentiment Classification.} Imdb Movie Review (\imdb)~\citep{maas-EtAl:2011:ACL-HLT2011} is a large-scale movie review dataset for sentiment analysis. Stanford Sentiment (\sst)~\citep{socher2013recursive} is another movie review dataset with relatively shorter text than \imdb. Financial Sentiment (\fst)~\citep{malo2014good} provides sentiment labels on economic texts in finance domain.  We use \imdb to train local victim models and consider \sst and \fst as target domains in attack.

\paragraph{Machine Translation.} We consider German (De) to English (En) translation as our testbed. We first study the attack performance on the local models trained on a general domain. Specifically, we use WMT14~\citep{bojar-EtAl:2014:W14-33} to train the victim models. Then, we investigate the imitation attack on \textbf{Law} and \textbf{Koran} domains from OPUS~\citep{tiedemann2012parallel}. We utilize Moses\footnote{https://github.com/moses-smt/mosesdecoder} to pre-process all corpora, and keep the text cased. A 32K BPE vocabulary~\citep{sennrich2016neural} is applied to all datasets. 

\subsection{Victim Models}

\paragraph{Locally Simulated NLP Services.} We first use models trained on our local server as local services. Our models are trained on datasets in source domain, \textit{i.e.}, \imdb for sentiment analysis and \textbf{WMT14} for machine translation. We vary BERT~\citep{devlin2019bert} and RoBERTa~\citep{liu2019roberta} as pre-trained models for classification. Transformer-base (TF-base) and Transformer-big (TF-big)~\citep{vaswani2017attention} are used as machine translation architectures, more details about hyper-parameter settings are described in Appendix~\ref{append:hyper_param}.

\paragraph{Commercial NLP API Services.}
To investigate the performance of imitation attack on real-world commercial NLP APIs, we query and retrieve the results of victim APIs for both sentiment analysis and machine translation. 
\googlesentiment API\footnote{https://cloud.google.com} and \ibmsentiment API\footnote{https://cloud.ibm.com} are inquired for sentiment analysis. \googletranslation\footnote{https://translate.google.com/} and \bingtranslation \footnote{https://www.bing.com/translator} are used as translation APIs. 
In this setting, we assume that different companies have various choices in datasets, domains, model architectures and training strategies. These settings are invisible to the attackers.

\begin{table*}[t]
\begin{center}
\begin{tabular}{ l llrr llrr } 
 \toprule
  & \multicolumn{4}{c}{\textbf{Sentiment Classification}} & \multicolumn{4}{c}{\textbf{Machine Translation}}\\
 \cmidrule(lr){2-5} \cmidrule(lr){6-9} 
 \textbf{Model} & \textbf{Label} & \textbf{Arch.} &\textbf{SST}  & \textbf{FST} & \textbf{Label} & \textbf{Arch.} & \textbf{Law}  & \textbf{Koran}\\
 \midrule
 \midrule
 In Domain & SST / FST & BERT & \color{gray}{91.49} & \color{gray}{97.72}     & Law / Koran & TF base & \color{gray}{38.52} & \color{gray}{19.49}\\
 \midrule
 Victim 1 & IMDB & BERT & 87.92 & 74.94         & WMT14 & TF-base & 23.33 & 9.82\\
 Victim 2 & IMDB & RoBERTa & 89.40 & 80.00      & WMT14 & TF-big & 24.33 &  10.33\\
 \midrule
 Attack$_{s1}$ & Victim 1 & BERT & 90.13 & 83.59  & Victim 1 & TF-base & 23.82 & 10.04\\
 
  Attack$_{s2}$ & Victim 2 & BERT & 90.72 & 88.76 & Victim 2 & TF-base& 25.48 & 10.30\\
 \midrule
 Attack$_{m}$ & Victim 1+2 & BERT & \textbf{91.57} & \textbf{90.53} & Victim 1+2 & TF-base & \textbf{25.74} & \textbf{10.48}\\
 \bottomrule
\end{tabular}
\end{center}
\caption{Experimental results of imitation attack on single and multiple victim models with settings, labels used for training and model architectures (Arch.). Oracle models are trained on the human-annotated datasets in target domain and victim models are trained on corpora from source domain. For all attack experiments, we report the mean results over 5 runs. Attackers using single victim and multiple victims are indicated as Attack$_s$ and Attack$_m$. \textit{In domain} models are trained on human labeled training sets in the target domain, which could arguably be considered as upper bounds of those attack models. This experiment focuses on comparing Victim and Attack.}
\label{tab:local_sent_performance}
\end{table*}

\subsection{Imitation Attack Setup}
For imitation attack, different from~\citet{wallace2020imitation}, we leverage datasets in other domains rather than those used for training victim models. The rationale behind this setting is that \textit{i)} the owners of APIs tend to use in-house dataset, which is difficult for attacker to access; \textit{ii)} attackers are more interested in their own private dataset, which is also not visible to others. Therefore, our setting is closer to the real-world attack scenario.
The attack models are trained based on the labels retrieved from victim models, and tested on the labels from the human-annotated sets.
We consider \sst and \fst as target domains for sentiment analysis.
For machine translation, we use \textbf{Law} and \textbf{Koran}. In attack, we use BERT for classification. Regarding machine translation, Transformer-base is used for simulating local APIs, while mBART~\citep{liu2020multilingual} for experiments on commercial APIs\footnote{We observe that mBART works better on attacking commercial APIs than Transformers in our preliminary experiments.}.
We also investigate the ensemble models, by concatenating all the outputs retrieved from multiple victim models in training.



\section{Experimental Results}
\label{sec:exp_result}

In this section, we analyze our experimental results. Our experiments are designed to answer the following research questions (\textbf{RQ}s),
\begin{itemize} 
    \item \textbf{RQ1}: Are the attack models able to outperform the victim models in new domains?
    \item \textbf{RQ2}: Will the ensemble of victim APIs improve the performance of the attack models?
    \item \textbf{RQ3}: Do traditional defense methods help APIs reduce the performance of attackers in our domain adaptation setting?
\end{itemize}

\textbf{Locally Simulated Experiments}. We first conduct the experiments of imitation attack on local models, shown in Table~\ref{tab:local_sent_performance}. The models trained on oracle human annotated datasets are much better than the victim models, as the later ones are trained on other domains. All our attack models outperform the original victim models for both classification and translation tasks. We attribute this performance improvement to unsupervised domain adaptation. Ensemble models consistently work better than each of the single model\footnote{More discussion on averaging strategies for ensemble the classification APIs is discussed in Appendix~\ref{append:ensemble_comparison}.}. For SST, although using the same architecture, the attack model trained on the ensemble of two victims surprisingly outperforms the model supervised by oracle labels. This result also outperforms some competitive supervised baselines~\citep{tang2019distilling,mccann2017learned,zhou2016text}. This observation suggests that, in some scenarios, it is possible to achieve decent results only based on some open APIs, without relatively more expensive human annotations. 
As a result, some annotators could lose their working opportunities of labeling new datasets, and some API services might lose their market share in new domains or tasks.

\begin{table*}[t]
\begin{center}
\begin{tabular}{ ll|ll ll r } 
 \toprule
 \textbf{Task} & \textbf{\# Queries} & \textbf{API} & \textbf{Cost} & \textbf{Victim} & \textbf{Attacker} & Improv.\\
 \midrule\midrule
 \sst & 9,613 & \googlesentiment & \$5 & 84.62 & 88.26 $\pm$ 0.22 & +3.64\\
      & & \ibmsentiment & Free & 87.26 & 89.17 $\pm$ 0.33 & +1.91\\
      \cline{5-7} 
      & & \textbf{Google}+\textbf{IBM} & \$5 & - & 89.75 $\pm$ 0.58 &\\
 \midrule
 \fst & 1,968 & \googlesentiment & Free & 68.35 & 83.85 $\pm$ 1.05 & +15.50\\
      & & \ibmsentiment & Free & 58.73 & 85.01 $\pm$ 0.81 & +26.28\\
      \cline{5-7} 
      & & \textbf{Google}+\textbf{IBM} & Free & - &  89.82 $\pm$ 0.81 \\
     \midrule
 \textbf{Law} & 2M & \googletranslation & \$6,822 & 30.43 & 31.99 $\pm$ 0.05 & +1.56\\
      & & \bingtranslation & \$3,396 & 34.22 & 35.45 $\pm$ 0.09 & +1.23\\
      \cline{5-7} 
      & & \textbf{Google}+\textbf{Bing} & \$10,218 & - & 34.94 $\pm$ 0.11\\
\midrule
      \textbf{Koran} & 579k & \googletranslation & \$1,211 & 14.31 & 14.63 $\pm$ 0.06 & +0.32\\
      & & \bingtranslation & \$590 & 13.24 & 13.71 $\pm$ 0.05 & +0.47\\
      \cline{5-7} 
      & & \textbf{Google}+\textbf{Bing} & \$1,801 & - & 15.25 $\pm$ 0.09 \\
 \bottomrule
\end{tabular}
\end{center}
\vspace{-2mm}
\caption{A comparison of the commercial APIs (Victims) with attackers. The improvement (Improv.) of attackers over victims is given by rows of single models. The estimated cost of the API is based on the price queried in a single day. For all attack experiments, we report mean and standard deviations of the results over 5 runs.}
\vspace{-2mm}
\label{tab:API_performance}
\end{table*}

\textbf{Experiments on Commercial APIs}. We then demonstrate the vulnerability of some real-world commercial APIs to our IMA approach, in Table~\ref{tab:API_performance}. For classification task, the attacker uses soft labels, as \textit{i)} these APIs provide such scores and \textit{ii)} attackers using soft labels achieve better performance than hard attacks in our defense experiments (see Table~\ref{tab:defense_strategies}). For machine translation, only hard label could be used, as we can only have token sequences without their perplexity scores from commercial APIs. 
In all attacks on classification and translation APIs, the attackers manage to achieve significantly better results than the corresponding victim models, with frighteningly low costs. 
Combining two commercial APIs generally improves the performance of the attackers.
We observe that commercial APIs work quite poor on FST, as it 
belongs to a more professional domain. However, the performance of the attacker catches up significantly on FST and achieves an averaged accuracy of 89.82\%, given poor competitors (victims) both with accuracy less than 70\%.
Google+Bing on Law is the only ensemble model that fails to surpass all the single model. We attribute this to the fact that Bing Trans and its attack model has already achieved a decent result on Law, outperforming Google Trans with a gap of 3\% to 4\%.

\textbf{Estimated Attack Costs}. In Table~\ref{tab:API_performance}, we also estimate the cost of querying the commercial APIs as victim models. We find the costs are quite affordable to many companies or even individuals, given the benefit of obtaining high-quality in-domain classification and MT systems. The price is in accordance with retrieving the results in a day, using a single month budget of a single account. The price could be further decreased, by registering more accounts or using more time. On the other hand, we estimate that the costs for human to annotate the datasets are \$480.65 (\sst), \$98.4 (\fst), \$1.6M (\textbf{Law}), and \$463k (\textbf{Koran}), if we hire annotators from Amazon Mechanical Turk. The price is decided as 0.05 cents for each classification sample and 0.8 dollar for each translation sample\footnote{In our preliminary experiment, annotators manage to finish about 10 classification and 1.5 translation annotations. The corresponding wages are about \$30/hr and \$32/hr, higher than the minimum wage in USA. }. Although the price is arguably floating, we give a preliminary overview on the costs by human annotators. We find that the cost of human annotation could be 
20 to 150 times higher than querying APIs. This could become another motivation for attackers to learn from APIs, instead of human annotators.

\begin{table}[t]
    \centering
    \scalebox{0.77}{
    \begin{tabular}{l c ccc ccc}
    \toprule
    \multicolumn{2}{c}{} & \multicolumn{2}{c}{\textbf{SST}} & \multicolumn{2}{c}{\textbf{Law}}\\
    \cmidrule(lr){3-4} \cmidrule(lr){5-6} 
    Model  & \#V & Victim & Attack & Victim & Attack \\
    \midrule \midrule
    Model A & 1 & 85.39 & 87.37 & 20.75 & 22.60\\
    Model B & 1 & 84.51 & 86.22 & 21.01& 21.67\\
    Model C & 1 & 86.44 & 87.70 & 20.79 &21.76\\
    Model D & 1 & 86.60 & 87.31 & 20.68 &21.80\\
    \midrule
    Model A+B     & 2 & - & 88.30 &- & 22.87\\
    Model A+B+C+D & 4 & - & 89.18 & - &22.97\\
    \midrule
    Model Full & 1 & 87.92 & 88.58 & 23.33& 23.82\\
    \bottomrule
    \end{tabular}
    }
\vspace{-4mm}
\caption{The comparison of imitation attacks with different model ensembles. 
}
\label{tab:multi_victim_ablation}
\vspace{-2mm}
\end{table}
\textbf{Impact of Model Ensemble}. We conduct an ablation study on the potential performance improvement of attacking multiple victim models, in Table~\ref{tab:multi_victim_ablation}. The source domain training datasets are equally separated into 4 disjoint subsets, coined A, B, C and D. Then, we train 4 corresponding local victim models, model A to D, respectively. Model Full uses the victim models trained on the complete training datasets in the source domains.
The attacker chooses to utilize the combined results of these victim models for training, \textit{e.g.}, Model A+B ensembles knowledge from victim A and B. We use BERT for classification and Transformer-base for machine translation in this study. Ensemble models with more victims generally improve the attacker performance on SST and Law. The ensemble of 
multiple weaker models can catch up with those of full model. 

\begin{table*}[t]
    \centering
    \begin{tabular}{l || c|c| ccc || c|c| ccc}
    \toprule
    & \multicolumn{5}{c||}{SST} & \multicolumn{5}{c}{FST}\\
    \cline{2-11}
    Model & Soft & Hard & P 0.1 & P 0.2 & P 0.5 & Soft & Hard & P 0.1 & P 0.2 & P 0.5\\
    \hline
    Victim 1 & 87.92 & 87.92 & 87.94 & 86.79 & 78.06 & 74.94 & 74.94 & 75.24 & 69.27 &58.33  \\
    Victim 2 & 89.40 & 89.40 & 88.79 & 87.73 & 80.71 & 80.00 & 80.00 & 78.08 & 76.46 &65.42  \\
    \hline
    Attack$_{s1}$ & 90.44 & 88.58 & 90.23 & 90.07 &87.98 & 82.03 & 80.25 & 83.65 & 83.75 &76.41 \\
    Attack$_{s2}$ & 90.12 & 90.12 & 90.49 & 90.47 &88.72  & 87.85 & 85.57 & 88.86 &85.16 &82.03\\
    Attack$_{m}$ & 91.82 & 90.66 & 91.44 & 91.20 &90.15 & 88.86 & 87.09 & 89.27 &87.34 &86.33 \\
    \bottomrule
    \end{tabular}

\caption{The comparison of imitation attack results given victims with various defense strategies, soft label (Soft) to hard label (Hard) and noise perturbation (P) with variance ($\sigma$).}
\vspace{-4mm}
\label{tab:defense_strategies}
\end{table*}

\textbf{Defense Strategies}.

We compare two possible defense strategies on classification tasks in Table~\ref{tab:defense_strategies}. We consider perturbing the original soft outputs by hard labeling or adding Gaussian noise $\vn \sim \mathcal{N}(0, \sigma^2)$, as $$ h'_v(x) = h_v(x) + \vn.$$ The models utilizing hard labels manage to consistently reduce the performance of the attackers who use soft labels. Then we compare models trained on labels with various perturbation (P) by sampling random noise from Gaussian distribution with variance $\sigma\in \{0.1,0.2,0.5\}$~\citep{tramer2016stealing}. These experiments allow the noise to flip the outputs of the original victim models, and more experiments on perturbation without influencing utility of victim models are provided in Appendix~\ref{apped:defense_strategy}. We observe that gently disturbing the outputs of victim models could crack down the attackers to some extent and larger noise indicates better defense. However, the harm to victims is generally larger than those to attackers. We attribute this to noise reduction in training attack model. Overall, our new IMA calls for more effective defense methods in future.

\section{Discussion}
We consider our imitation attack approach has achieved outperforming results that challenge the current understanding of IMA. As a result, API publishing strategies and defense methodologies should be converted accordingly.




\textbf{Suggested Actions.}
As in domain adaptation settings, IMA manages to achieve superb performance, while attacking models are not able to outperform victims in the same domain~\citep{krishna2019thieves,wallace2020imitation}, we suggest API services could cover more domains to eliminate the potential performance gain from UDA.
Simply harming the utility of victim models seems not to be a wise choice for service providers, but merely providing hard labels without probability scores could avoid more superior attack performance.
Adjusting the pricing strategies of publishing API services or adding watermark to API outputs~\citep{he2021protecting} could be plausible solutions to avoid the illegal stealing of the precious APIs of industrial companies.

\textbf{Ethical Concerns.}
We recognize that our work could be used for malicious purposes, for example, a competing company may adopt our attack to steal a better model for commercial benefits, thus eroding the market shares of other business competitors. However, the main purpose of our work is to help commercial cloud services and regulators raise awareness of model theft, and reconsider how to deploy NLP APIs in a safe manner to avoid being stolen and exceeded. 
In order to minimize the potential negative influence of our work, \textit{i)} we will delete our models and retrieved results on our local server; \textit{ii)} we will report the vulnerability to the API providers.

\textbf{Follow-up Attacks}. Recent works have demonstrated that the extracted model could be used as a reconnaissance step to facilitate later attacks~\citep{he2021model,krishna2019thieves,wallace2020imitation}. For instance, the adversary could use the extracted model to 
facilitate private information inference about the training data of the victim model, or to
construct adversarial examples that force the victim model to make incorrect predictions. We leave follow-up attacks that can leverage our imitated model with better performance to future works.

\section{Conclusion}
\label{sec:Conclusion}
We demonstrate a powerful imitation attack paradigm which may produce better attack models that surpasses the imitated models including the real-world NLP APIs via domain adaptation and ensemble. We believe such achievements would potentially influence the price decision and publishing strategies of primary NLP services.
We also take the first step of grounding our new attacking approach with unsupervised domain adaptation theory and model ensemble.
More broadly, we hope to arouse prominent concerns on security and privacy of API services in NLP applications. In future, we plan to further explore the possibility of detecting and defending the imitation attack.


\bibliography{custom}
\bibliographystyle{acl_natbib}

\clearpage
\appendix

\clearpage
\onecolumn

\section{Hyper-Parameter Settings}
\label{append:hyper_param}
In order to have a fair and consistent comparison of experiments, we utilize the same hyper-parameters for the same task, as demonstrated in Table~\ref{tab:hyper}. They are decided by our preliminary experiments on target domains. All experiments could be conducted on a single RTX 6000 GPU.

\begin{table}[h]
    \centering
    \begin{tabular}{l c c}
    \toprule
     & Sent. & MT \\
    \midrule
    Learning rate & 1e-05 & 5e-04\\
    Batch size & 16 sentences & 32k tokens\\
    Optimizer & Adam & Adam\\
    Epoch & 50 & 40\\
    Max length & 256 & 1024\\
    Warm-up & - & 4000 steps \\
    \bottomrule
 \end{tabular}
    \caption{Hyper-parameter used for sentiment analysis (Sent.) and machine translation (MT).}
    \label{tab:hyper}
\end{table}

\section{Comparison of Ensemble Strategy}
\label{append:ensemble_comparison}

In this section, we compare two ensemble methods for sentiment classification, i) concatenating the training samples (Concat.) and ii) averaging the prediction scores (Avg.). Both ensemble strategies are competitive to the other, as demonstrated in Table~\ref{tab:ensemble_avg_concat}. 
As we are not able to acquire scores of each token from MT APIs, we cannot average the results of MT. In our paper, to have a consistent comparison, Concate. is used in all our ensemble experiments on both classification and translation tasks.


\begin{table}[h]
    \centering
    \begin{subtable}{.49\textwidth}
    \centering
    \begin{tabular}{c c  c}
    \toprule
    Method & Ensemble & Accuracy \\
    \midrule
    BERT+RoBERTa & Concat. & 91.57 $\pm$ 0.27\\
    BERT+RoBERTa & Avg. & 91.58 $\pm$ 0.39 \\
    Google+IBM & Concat. & 89.75 $\pm$ 0.58\\
    Google+IBM & Avg. & 89.76 $\pm$ 0.42\\
    \bottomrule
    \end{tabular}
    \caption{Ensemble results on SST.}
    \end{subtable}
~
    \begin{subtable}{.49\textwidth}
    \centering
    \begin{tabular}{c c  c}
    \toprule
    Method & Ensemble & Accuracy \\
    \midrule
    BERT+RoBERTa & Concat. & 90.53 $\pm$ 0.38\\
    BERT+RoBERTa & Avg. & 90.89 $\pm$ 0.48\\
    Google+IBM & Concat. & 89.82 $\pm$ 0.81\\
    Google+IBM & Avg. & 88.86 $ \pm$ 1.22\\
    \bottomrule
    \end{tabular}
    \caption{Ensemble results on FST.}
    \end{subtable}
    \caption{The comparison of imitation attack on multiple victims using concatenate samples (Concat.) and average scores (Avg.).}
    \label{tab:ensemble_avg_concat}
\end{table}

\section{Comparison of Defense Strategies}
\label{apped:defense_strategy}
In this section, we discuss the perturbation methods. Given an input sentence $x$, the probability score by the victim model is $h_v(x)$. To compare the influence of API performance on attack model, we sample a noise vector $\vn$ from Gaussian distribution with a variance of $\sigma$, \textit{i.e.}, $\vn \sim \mathcal{N}(0, \sigma^2)$. The perturbed prediction function is calculated as:
\begin{equation*}
    h'_v(x) = h_v(x) + \vn
\end{equation*}

\begin{table*}[h]
    \centering
    \begin{tabular}{l || c|c| ccc || c|c| ccc}
    \toprule
    & \multicolumn{5}{c||}{SST} & \multicolumn{5}{c}{FST}\\
    \cline{2-11}
    Model & Soft & Hard & P 0.1 & P 0.2 & P 0.5 & Soft & Hard & P 0.1 & P 0.2 & P 0.5\\
    \hline
    Victim 1 & \multicolumn{5}{c||}{87.92}  & \multicolumn{5}{c}{74.94} \\
    Victim 2 & \multicolumn{5}{c||}{89.40} & \multicolumn{5}{c}{80.00} \\
    \hline
    Attack$_{s1}$ & 90.44 & 88.58 & 90.48 & 89.96 & 89.35 & 82.03 & 80.25 & 82.84 & 81.92 & 81.72 \\
    Attack$_{s2}$ & 90.12 & 90.12 & 90.30 & 90.02 & 89.23 & 87.85 & 85.57 & 88.10 & 87.29 & 84.41 \\
    Attack$_{m}$ & 91.82 & 90.66 & 91.24 & 91.38 & 91.11 & 88.86 & 87.09 & 89.37 & 88.25 & 87.59 \\
    \bottomrule
    \end{tabular}
\caption{The comparison of imitation attack results given victims with various defense strategies, soft label (Soft) to hard label (Hard) and noise perturbation (P) with variance ($\sigma$). The predicted labels of victim models are not flipped in the experiment.}
\label{tab:defense_strategies2}
\end{table*}

It is worth noting that original victim model prediction is $y=\mathrm{argmax}(h_v(x))$, therefore, injecting $\vn$ could lead to different prediction $y'=\mathrm{argmax}(h'_v(x))$. Consequently, the utility of victim model can be corrupted as demonstrated in Table~\ref{tab:defense_strategies}. However, such compromise can cause finance and reputation losses to the API providers in real-world. To avoid these adverse effects, API providers can adopt a label-preserved policy, where the injected noise $\vn$ should not flip the originally predicted labels. In other words, another noise will be sampled, if currently sampled noise changes the prediction of the original model. The results of such defense strategy is shown in Table~\ref{tab:defense_strategies2}. As this setting is more conservative, the performance of this defense is in between the performance of hard label and soft label.

\end{document}